\documentstyle[12pt]{article}
\newcommand{\simgeq}{\; \raisebox{-0.4ex}{\tiny$\stackrel
{{\textstyle>}}{\sim}$}\;}

\newcommand{\beq}{\begin{equation}}
\newcommand{\beqar}{\begin{eqnarray}}
\newcommand{\eeq}[1]{\label{#1} \end{equation}}
\newcommand{\eeqar}[1]{\label{#1} \end{eqnarray}}
\newcommand{\eps}{\varepsilon}
\input{psfig}

\begin{document}

\vspace*{2cm}
\begin{center}
{\Large{\bf Triaxiality in $^{48}$Cr}}
\vspace{1cm}

{\bf Andrius Juodagalvis%
\footnote{\ Tel: +46-46-222 9087, fax: +46-46-222 4416, 
e-mail: AndriusJ@MatFys.LTH.se}%
, Ingemar Ragnarsson and Sven \AA berg}
\\
\begin{footnotesize}
Div.\ of Mathematical Physics, Lund Institute of Technology, P.O.\ Box
118, S-221 00 Lund, Sweden
\\
\smallskip
September 30, 1999\\
(Submitted to {\it Phys.Lett.B})
\end{footnotesize} 
\end{center}

\vspace{0.5cm}
\noindent
\begin{small}
{\bf Abstract:}
Rotational behavior inducing triaxiality is discussed for $^{48}$Cr
in the cranked Nilsson-Strutinsky (CNS) model, as well as in the spherical 
shell model. It is shown that the low-spin
region up to about $I$=8, has a prolate well-deformed shape. At higher
spins the shape is triaxial with a ``negative-$\gamma $'' deformation,
that is, with rotation around the classically forbidden
intermediate axis.
By comparing calculated $B(E2)$-values and spectroscopic quadrupole moments
in the CNS with
 spherical shell model results and experimental data, 
the triaxial rotation around the intermediate axis is confirmed.
\vspace{0.5cm}


\end{small}

\hspace{0.5cm}

Quantum triaxiality is a most interesting topic in nuclear structure
physics. Although the possibility to obtain nuclear shapes with all
three axes unequal has been discussed for many years (see e.g.\
ref.\ \cite{Ha90}), many unanswered questions exist. For example, the
predicted wobbling mode \cite{BM2,An76} has not yet been observed in
any nucleus (although preliminary experimental information \cite{Ha99}
may indicate the existence of a wobbling band in $^{163}$Lu). In
classical mechanics rotation around the axis with the middle moment
of inertia is forbidden. In the (one-dimensional) cranking approach
this would correspond to ``negative-$\gamma$'' rotation (in the Lund
convention on $\gamma$ \cite{An76}), and it is still an open question
if such a coupling between the intrinsic axes and the rotational axis is
realized. Electromagnetic properties, such as quadrupole transitional
($\Delta I$=2 and 1) and spectroscopic moments, as well as $M1$ and
$g$-factors, are expected to be strongly dependent on the
triaxiality. They also depend on the rotation axis, and may thus serve
as good tests of suggested coupling schemes, and on triaxiality
\cite{Ha90}.

In Ref.\ \cite{An76} it was concluded that particles in high-$j$ shells
which are half-filled or close to half-filled tend to polarize the
nucleus towards negative $\gamma$-values. From this point of view,
$^{48}$Cr should be a good candidate for this kind of rotation
because both the proton and the neutron configurations correspond
to half-filled $f_{7/2}$-shells. 

The nucleus $^{48}$Cr is well studied experimentally \cite{Cr48}, as
well as in extensive shell model calculations
\cite{Poves-Cr,Poves-A48}. By comparing mean field cranked
Nilsson-Strutinsky (CNS) calculations with full $fp$-shell model
calculations, conclusions can be drawn about the rotational behavior
as a function of angular momentum. From calculated $B(E2)$ values and
spectroscopic quadrupole moments we find that $^{48}$Cr indeed becomes
triaxial at higher spins, with rotation around the intermediate axis
(``negative-$\gamma$ rotation'').

\begin{figure}[tbp]
\centerline{\psfig{figure=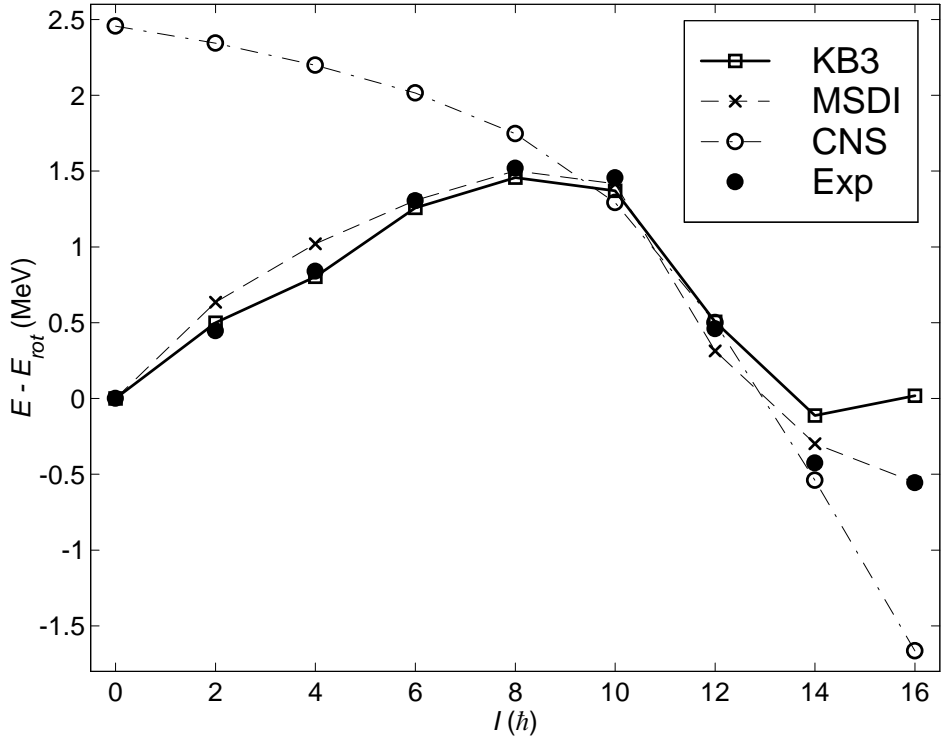,width=0.8\textwidth}}
\caption{Calculated and measured energies for yrast states
in $^{48}$Cr. The solid line shows values calculated in the shell model
with the KB3-interaction \cite{Poves-A48} (model space is full $fp$ shell), 
and the dashed
line shows shell model results obtained with the modified 
surface-delta interaction, MSDI \cite{And} (model space is $f_{7/2}p_{3/2}$).
The dot-dashed line shows the calculated energies in the cranked 
Nilsson-Strutinsky model without pairing, CNS.
Measured energies are shown by filled circles \cite{Cr48}.
A rotational reference, 
$E_{rot}$=0.051\,$I(I+1)$, has been subtracted.
}
\label{fig1}
\end{figure}

Measured
and calculated energies of the ground-state band of $^{48}$Cr are shown
in fig.~\ref{fig1}. For 
convenience a rotational reference has been subtracted. %
It is seen how the rotational behavior, the backbending and band
termination, are all very well described by the shell model
calculations, while the CNS calculation is not able to describe
neither the low-spin regime nor the transition energy of the
terminating $16^+$ state.  In the CNS calculations, the
configuration-constrained approach described in
Refs. \cite{Ragnarsson,Afa95} is used with standard single-particle
parameters \cite{Ragnarsson} for the Nilsson potential.  In the
ground-state, the 8 valence particles occupy the (deformed) $f_{7/2}$
shell. This corresponds to positive parity and signature 0 (even
spins), and gives lowest energy for all states up to 16$^+$. We allow
for a free minimization in the two quadrupole degrees of freedom,
$\eps$ and $\gamma$, as well as in one hexadecapole degree of freedom,
$\eps_4$.  The discrepancy between calculation and experiment at low
spins is expected since no pairing is included in the CNS calculation.
The underestimation by about 1 MeV in the description of the band
terminating state at $I^{\pi}$=$16^+$, indicates that pairing (mainly
proton-neutron $T$=1 pairing \cite{And}) plays an important role also
at quite high spin states.

The shell model results show the outcome of two calculations, one
performed in the full $fp$ shell with the KB3 interaction (cf.\
\cite{Poves-A48}), and another performed in the restricted subspace
$f_{7/2}p_{3/2}$ with the modified surface-delta interaction, MSDI
(cf.\ \cite{And}). When calculating electromagnetic properties, in both
cases standard effective charges $q_\pi$=1.5$e$ and $q_\nu$=0.5$e$
were used.
The calculations were performed using code ANTOINE
\cite{Caurier-code}. Notice that the shell model calculation in the
restricted subspace $f_{7/2}p_{3/2}$ (``quasi-SU(3)'' \cite{Zuker})
describes the energies equally well as the full $fp$ shell
calculation. The distribution of shell occupancy in $f_{7/2}$ is
actually quite similar, not only in the two shell model calculations,
but also in the CNS calculation
\cite{An99}, as well as in CHFB \cite{Poves-Cr},
in spite of the fact that the energies are poorly
described in the latter models.  This is a result of the quadrupole
deformation, that simulates the shell mixing very well \cite{Zuker}.
It is then interesting to study the properties that are specificly
dependent on the quadrupole degree of freedom, i.e.\ $E2$ transition
probabilities and spectroscopic quadrupole moments.

In fig.2 we show calculated and experimental $B(E2)$ values for
stretched quadrupole transitions along the ground-state band in
$^{48}$Cr.  It is seen that the calculated values in the full $fp$
shell (KB3 interaction), as well as in the restricted $f_{7/2}p_{3/2}$
subspace (MSDI interaction) come close to the measured $B(E2)$
values. Since the full $fp$ shell calculation agrees somewhat better
with the data we shall prefer this model in later
comparisons. However, the same conclusions can be drawn from the MSDI
calculation.
\begin{figure}[tbp]
\centerline{\psfig{figure=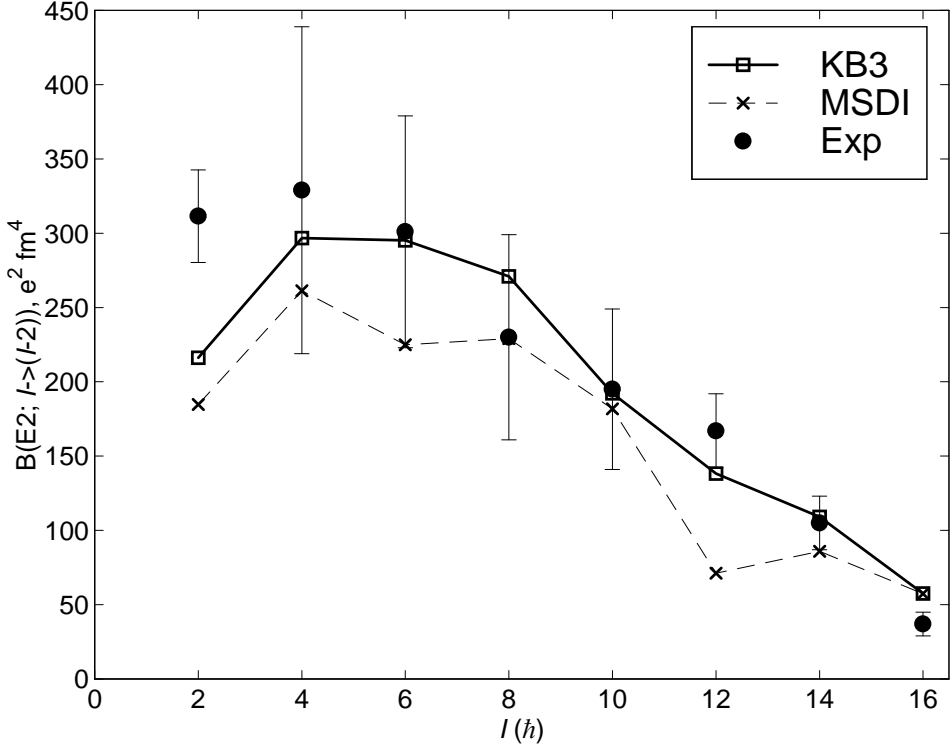,width=0.8\textwidth}}
\caption{Calculated and measured $B(E2)$-values for yrast states
in $^{48}$Cr. The solid line shows values calculated in the shell model
(full $fp$-shell) using the KB3-interaction \cite{Poves-A48}, and the dashed line 
shows calculated results in the restricted model space ($f_{7/2}p_{3/2}$)
using the modified surface-delta interaction. 
Measured $B(E2)$-values \cite{Cr48,Cr48-BE2} are shown with error bars.}
\label{fig2}
\end{figure}

The energies calculated in the CNS approach are illustrated above (fig.1), and
potential-energy surfaces for the states 0$^+$,  8$^+$, 12$^+$, and 14$^+$
are shown in fig.3. Resulting equilibrium deformations are shown in fig.4.
\begin{figure}[tbp]
\centerline{\psfig{figure=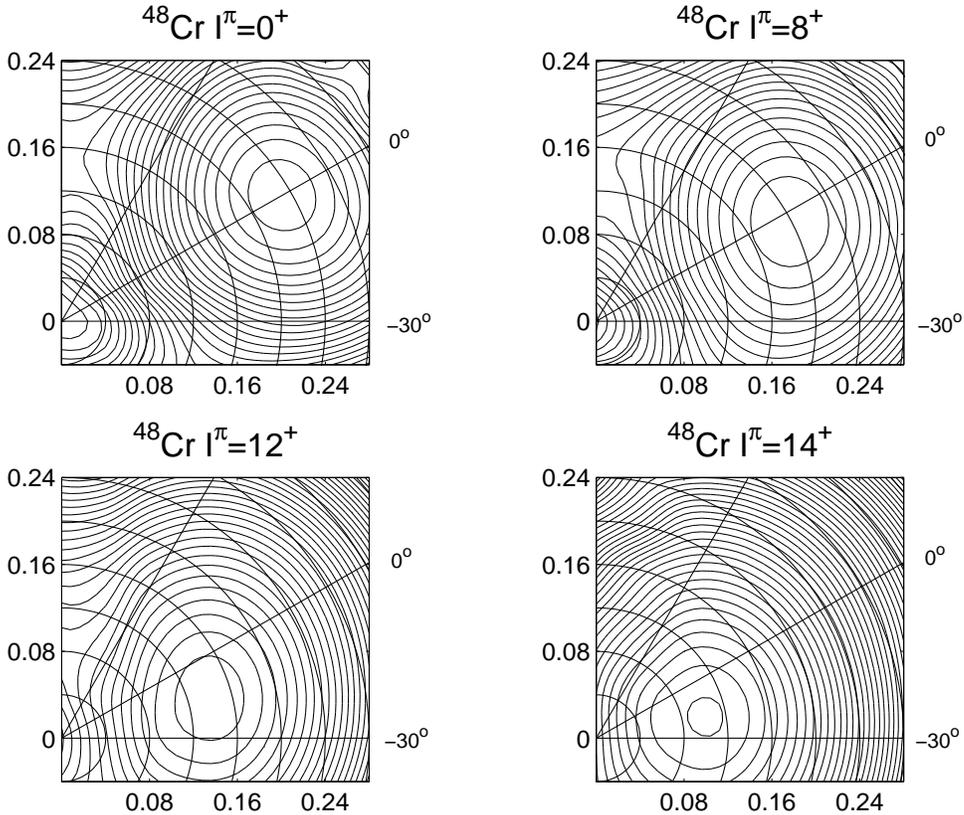,width=0.8\textwidth}}
\caption{Calculated potential-energy surfaces in the CNS model
for $^{48}$Cr at $%
I^{\pi}$=0$^+$, 8$^+$, 12$^+$, and 14$^+$. The line $\gamma$=$0^{\circ}$ 
corresponds to prolate shapes with collective rotation, and
$\gamma$=$60^{\circ}$ to oblate shapes and non-collective rotation. 
The contour line separation is 0.2 MeV.}
\label{fig3}
\end{figure}
\begin{figure}[tb]
\centerline{\psfig{figure=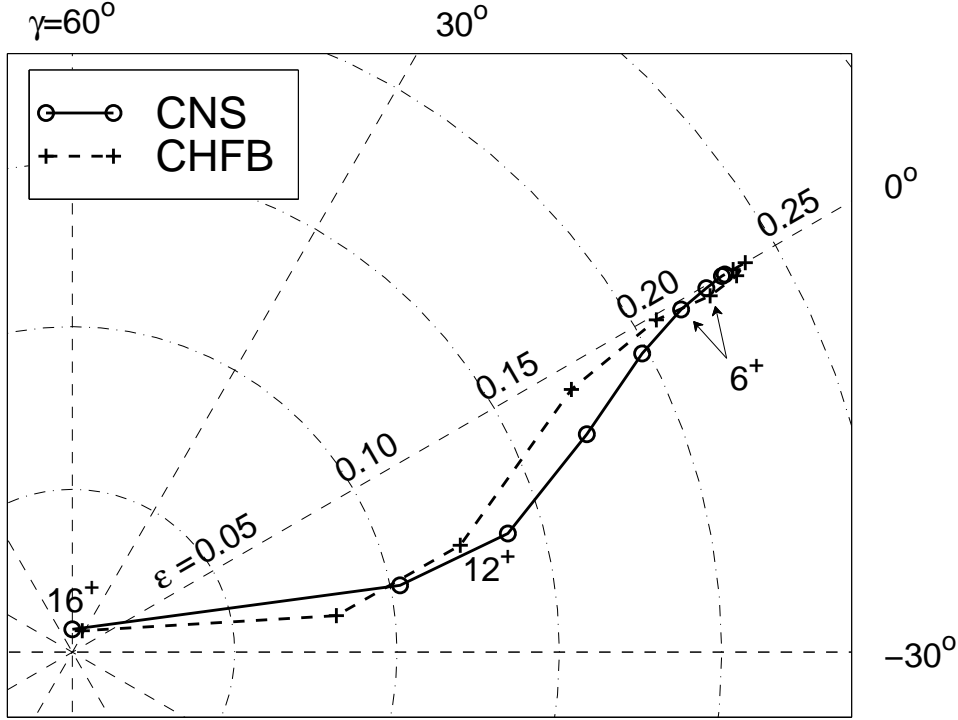,width=0.8\textwidth}}
\caption{Calculated equilibrium deformations in the 
($\eps$,$\gamma$)-plane for the yrast states, $I^{\pi}$=$0^+$-$16^+$,
in $^{48}$Cr. CNS values are shown by circles, while CHFB results \cite{Poves-Cr}
are shown by plus signs. The latter values were obtained transforming 
$(\beta,\gamma_\beta)$ to $(\varepsilon,\gamma)$ under the assumption that the 
ratio of axes should be preserved.}
\label{fig4}
\end{figure}
\noindent
The deformation of the ground state is found to be prolate
with $\eps$$\approx$0.26 with the four protons and four neutrons 
in the $f_{7/2}$ shell occupying the 
deformed orbits having $\Omega$=$\pm$1/2 and $\pm 3/2$.
The deformation shrinks somewhat as the angular momentum
increases, and the $8^+$ state has $\eps$$\approx$0.20\@. With
increasing angular momentum the equilibrium deformation clearly becomes
triaxial, and e.g.\ the $12^+$ state has $\gamma$$\approx$$-15^{\circ}$. This
corresponds to rotation around the (classically forbidden) intermediate 
axis. 
Finally, the 16$^+$ state is constructed by
aligning all 4 protons and 4 neutrons in the $f_{7/2}$ shell in the orbits
quantized along the rotation axis (the $x$-axis)
with $m_j$=+1/2, +3/2, +5/2 and +7/2, i.e.\ $I$=2$\cdot$8=16,
the maximal angular momentum
that can be built within this configuration. Thus, the rotational band 
terminates at 16$^+$ in an approximately spherically symmetric state.
Similar deformations are found in cranked Hartree-Fock-Bogoliubov (CHFB)
calculations with the finite range density dependent Gogny force 
\cite{Poves-Cr}, see fig.4.

Now two questions emerge: Does the 
CNS result for the quadrupole shapes have any physical 
relevance, in spite of the
rather poor description of energies (see fig.1), and if so,
can the suggested triaxial behavior with negative $\gamma$, be tested?
We shall approach the first question in a pragmatic way, namely
by calculating measurable properties in the CNS, and compare
to shell model results, which agree excellently with (available) 
experimental data.
This comparison will also show the relevance of the calculated 
triaxiality.

In the rotor model, assuming a fixed (axial) intrinsic electric
quadrupole moment of $Q_0(\hat{z})$, 
where $z$ is the symmetry axis,
stretched ($\Delta$I=2) $B(E2)$ values and spectroscopic quadrupole
moments $Q_{spec}$ are calculated as \cite[p.45]{BM2},
\begin{equation}
B(E2; I\!+\!2,K \rightarrow I,K)=
\frac{5}{16\pi}\langle I\!+\!2\, K 2 0 | I K\rangle ^2
Q_0(\hat{z})^2,
\end{equation}
and
\begin{equation}
Q_{spec}=\langle I I 2 0 | I I\rangle \langle I K 2 0 | I K\rangle 
Q_0(\hat{z}),
\end{equation}
respectively.
The motion is quantized along the $z$-axis, and $K$ is the angular
momentum component along this axis.
For triaxial shapes the $K$ quantum numbers are mixed and the above
relations cannot be used. If, however, $I$$\gg$$K$ it is possible
to derive similar expressions by 
instead quantizing the angular momentum along
the rotation axis, the $x$-axis.
For arbitrary triaxial shapes
the corresponding expressions become \cite[p.193]{BM2},
\begin{equation}
B(E2; I\!+\!2,K_x\!=\!I\!+\!2 \rightarrow I, K_x\!=\!I)=\frac{5}{16\pi}Q_2(\hat{x})^2,
\end{equation}
and 
\begin{equation}
Q_{spec}(I,K_x\!=\!I)=Q_0(\hat{x}),
\end{equation}
where $Q_2(\hat{x})$ and $Q_0(\hat{x})$ are the electric quadrupole moments
around the rotation axis ($x$-axis). The electric quadrupole moments
are calculated using the proton wave-functions at the appropriate 
equilibrium deformations.
In the case of axial symmetry and $\gamma$=$0^\circ$, 
$Q_2(\hat{x})$=$-\sqrt{3/8}Q_0(\hat{z})$ and
$Q_0(\hat{x})$=$-(1/2)Q_0(\hat{z})$, and the expressions in eqs.(3) and (4) 
coincide
with the high-spin limits of the expressions given in eqs.(1) and (2).
For the calculated low-spin shapes in $^{48}$Cr (axially symmetric) 
eqs.(1) and (2) may be applied, and for the higher spins 
($I\!\simgeq\!10$), the expressions (3) and (4) are 
reasonable approximations. To involve both regions
in one smooth expression, we combine the two formulae into,

\begin{equation}
B(E2; I\!+\!2,K \rightarrow I, K)=\frac{5}{16\pi}
\langle I\!+\!2\, K 2 0 | I K\rangle ^2 Q_2(\hat{x})^2 \cdot \frac{8}{3},
\end{equation}
and
\begin{equation}
Q_{spec}(I, K)=\langle I I 2 0 | I I\rangle \langle I K 2 0 | I K\rangle 
Q_0(\hat{x}) \cdot (-2),
\end{equation}
where the expressions have been divided by the asymptotic values for the 
Clebsch-Gordon (C-G)
coefficients ($(\sqrt{3/8})^2$, and $(-1/2)$, respectively).
These expressions are thus valid {\em either} at axial-symmetric shapes for
any $I$ and $K$, {\em or} at triaxial shapes for high-spin values 
($I$$\gg$$K$),
when the C-G coefficients have taken their asymptotic values. 
The expressions are not valid for triaxial shapes at lower spins.
In our study we obtain prolate shapes at low spins and triaxial
shapes at higher spins,
and we shall use the expressions as convenient interpolations
between the regions. The $K$ value in the above formula are then taken
from the axial-symmetric limit, i.e.\ $K$=0 for the ground band in $^{48}$Cr.

As one additional approximation, we shall neglect the change of
deformation between the mother state and daughter state in calculating
transitional quadrupole moments, and insert the wave function of the
mother state in calculating the $B(E2)$ values with the above
expressions. In the studied cases this gives smaller
$B(E2)$-values than the alternative method to use wave-functions of
daughter states, or states interpolated between mother and daughter
states.  We believe this approach is reasonable, since the
decay of the mother state is connected with a deformation change, that
should slow down the transition.

In principle, the shell model contains all types of correlations, 
corresponding to deformation degrees of freedom (including triaxiality),
three-dimensional rotation, quantum fluctuations, etc.
We therefore believe that by comparing calculated properties with 
shell model results, valuable information is obtained
about the validity and restriction of the cranking model, including
the above simplified expressions for $B(E2)$-values and spectroscopic
quadrupole moments.

In fig.5 we compare $B(E2)$ values 
calculated in the CNS, utilizing eq.(5), with those calculated in
the full $fp$ shell model, for the ground-state band in $^{48}$Cr. The
role of deformation changes is clearly seen. If the deformation is kept fixed
to its value in the 0$^+$ state, 
too large $B(E2)$ values result (short-dashed line in fig.5).
If the deformation is allowed to change, but is restricted to axial 
deformations, quite reasonable $B(E2)$ values are obtained; the values
are in fact similar to those obtained when allowing for full minimization
in the ($\eps$,$\gamma$,$\eps_4$) space. 
The effect of triaxiality with negative $\gamma$-values on $B(E2)$ values
is thus rather small, while if the calculated $\gamma$-values are used with
opposite signs (positive $\gamma$-values), too small $B(E2)$-values appear
(dot-dashed curve in fig.5).
\begin{figure}[tb]
\centerline{\psfig{figure=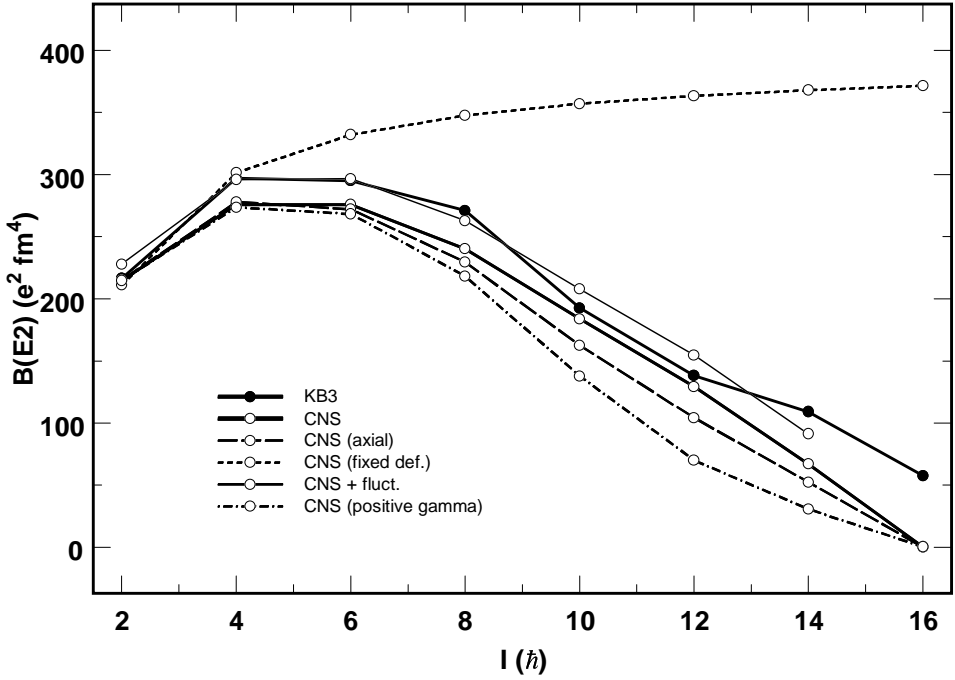,width=0.8\textwidth}}
\caption{Calculated $B(E2; I\! \rightarrow\! I\!-\!2)$-values 
for the yrast states of $^{48}$Cr. The curves show the
results obtained from the spherical shell model (full $fp$-shell) with the
KB3 interaction (filled circles connected by thick solid lines) 
and the CNS (empty circles connected by thick solid lines). 
The CNS results are also shown
assuming constant deformation (empty circles connected by short-dashed lines), 
deformations restricted
to axial shapes (empty circles connected by long-dashed lines), and triaxial
shapes with opposite signs on $\gamma$ (dot-dashed line).
Estimates of quantum fluctuations in the 
$\eps$-direction on the full CNS calculations are shown by empty circles
 connected by thin solid lines.}
\label{fig5}
\end{figure}

The CNS constitutes an approximate mean field theory, and quantum 
fluctuations may affect the results. In order to estimate the role
of quantum fluctuations
we perform a simple estimate of fluctuations in the
$\eps$ degree of freedom, that should play the dominant role 
for corrections on the $B(E2)$-values. 
If the equilibrium deformation at a given angular momentum state is
$\eps_0$, a harmonic approximation of the potential-energy
surface around this minimum yields,
\begin{equation}
E=E_0 + \frac{1}{2}C_{\eps}(\eps - \eps_0)^2,
\end{equation}
where the curvature, $C_{\eps}$, is obtained 
from the calculated potential-energy surfaces at each spin value, see fig.3.
The corresponding inertia parameter, $B_{\eps}$, is estimated
by \cite[p.531]{BM2}
\begin{equation}
B_{\eps}=12 B_{irr}.
\end{equation}
The zero-point motion in the $\eps$-direction gives an rms-value of the
quadrupole deformation,
\begin{equation}
\eps_{e\!f\!f}=\sqrt{\eps_0^2+\eps_{dyn}^2},
\end{equation}
where
\begin{equation}
\eps_{dyn}=\frac{1}{2}(B_{\eps}C_{\eps})^{-\frac{1}{2}}.
\end{equation}
Because the $B(E2)$ transition probabilities are proportional
to $\eps^2$,
we now estimate the enhancements on $B(E2)$-values due to
quantum fluctuations in the $\eps$-degree of freedom by simply 
calculating the electric quadrupole moment at the effective deformation,
$\eps_{e\!f\!f}$.
The results are shown in fig.5 by the thin solid line, and with
this estimate of the role of fluctuations the resulting $B(E2)$-values
are very similar to the shell model results (which include all kinds
of fluctuations).

Finally, in fig.6 spectroscopic quadrupole moments are shown. In the same
way as for the $B(E2)$ values the CNS results 
(calculated from eq.(6)) are shown under four
different assumptions, namely deformation fixed to the ground-state value,
free deformations but restricted to axial symmetry, free deformations
including triaxiality, and calculated triaxial shapes with opposite signs on
$\gamma$ (positive $\gamma$-rotation). 
Also here the importance of deformation changes is clearly seen. If the 
deformation is fixed to the ground-state shape
the spectroscopic moments are completely wrong.
However, most interesting is that the spectroscopic moments are 
indeed a good signature
for the proposed rotation scheme, namely that the nucleus obtains a triaxial
shape with rotation around the intermediate axis. For example the spectrocopic 
quadrupole moment of the 12$^+$ state
is calculated to be $-28$ e$\cdot$fm$^2$ if the shape is restricted 
to axial symmetry,
and $-14$ e$\cdot$fm$^2$ if the triaxial shape is considered, that 
is quite close to
the value obtained in the shell model calculation.

\begin{figure}[tb]
\centerline{\psfig{figure=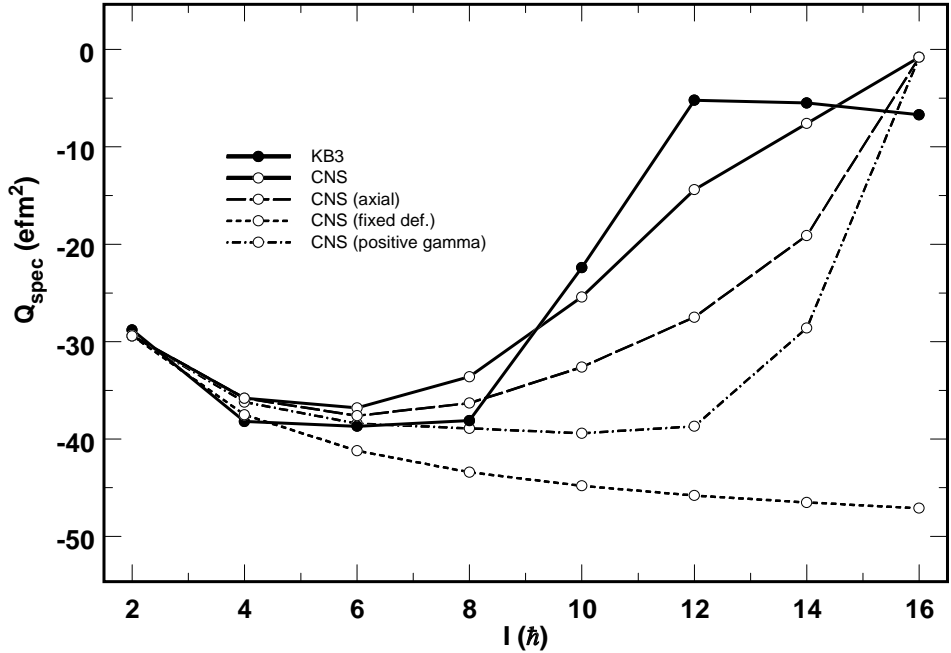,width=0.8\textwidth}}
\caption{Calculated spectroscopic quadrupole 
moments for the yrast states of $^{48}$Cr. Lines and symbols have
the same meaning as in fig.5}
\label{fig6}
\end{figure}

For rotation around the intermediate axis
quantum fluctuations in the $\gamma$-degree of freedom and in the
rotational axes (wobbling motion) are expected to be important in determining
spectroscopic quadrupole moments \cite{On86}
(because $Q_{spec}$$\propto$$\eps$, the
fluctuations in $\eps$ do not contribute to lowest order in this case).
The role of these fluctuations were found to be
particularly important at low spins 
and for small $\gamma$-values \cite{On86}
(cf.\ the discussion above about the validity of eq.(6)).
However, in our case the calculated equilibrium 
$\gamma$-deformations, that strongly affect the $Q_{spec}$-values,
are quite large and appear at substantial
spin values. This implies that our main 
conclusions are expected to remain also 
if these types of quantum fluctuations are included. Still, it would be
of interest to perform a fully dynamical calculation, i.e.\ to solve the
Bohr hamiltonian based on calculated high-spin
potential-energy surfaces for $^{48}$Cr.

With triaxial shapes existing over a few spin values a wobbling
excitation might principally be possible. We have 
looked for signatures of such an excitation mode
in the shell model calculations without success. The reason may be that
the triaxial deformation is not sufficiently stable and 
lasts over too few spin values. A more plausible explanation is,
however, that the relatively small number of
valence particles in $^{48}$Cr only allows for restricted collective 
phenomena.

In conclusion, we have studied the rotational behavior of $^{48}$Cr
and found that the CNS model describes the $B(E2)$ values as well as 
spectroscopic quadrupole moments very well, although this model
does not include all correlations (such as the pairing interaction) 
needed for an accurate description of the energy spectrum. 
The spin states above 
8$^+$ are found to be triaxial with rotation around the intermediate axis,
and the suggested rotation scheme is confirmed by the spherical shell model
calculations. It would be most
interesting if these results could be confirmed also by experimental data.

We would like to thank E. Caurier for access to the shell model code 
\cite{Caurier-code}.
A. J. thanks the Swedish Institute (``The Visby Programme'') for
financial support, and
I.R. and S.\AA. thank the Swedish Natural Science Research Council (NFR).

\end{document}